\documentstyle[11pt,newpasp,twoside,epsf]{article}
\def\edcomment#1{\iffalse\marginpar{\raggedright\sl#1\/}\else\relax\fi}
\reversemarginpar

\begin{document}
\title{Environment Shaping Galaxies: Star Formation Histories of Cluster Galaxies}
\author{Bianca Maria Poggianti}
\affil{INAF-Osservatorio Astronomico di Padova, vicolo dell'Osservatorio 5, 35142 Padova, Italy}

\begin{abstract}
In this paper I present some recent and new results regarding the effects
of the cluster environment on the star formation history of galaxies.
Three main aspects are discussed: the differences in the stellar
population ages of cluster ellipticals and S0 galaxies; the comparison
of the spectroscopic properties of galaxies in distant clusters
and in Coma; and the incidence and properties 
of faint post-starburst/post-starforming galaxies in the Coma cluster. 
\end{abstract}

\section{Introduction}
In the quest for undertanding how galaxy evolution is affected
by environmental processes, galaxy optical spectra provide 
precious informations regarding the evolutionary histories
and stellar population content of galaxies. Here I summarize 
some of the latest results obtained from spectroscopy of cluster
galaxies, both at low and high redshift.

\section{Lenticular galaxies}
Lenticular (S0) galaxies are a key component in clusters, being very
numerous and sometimes the most numerous of all Hubble types in rich
clusters today. Already more than 20 years ago it was suggested that
S0 galaxies could evolve from spirals that lost their gas supply.  The
hypothesis of a spiral origin for a significant fraction of the
cluster S0's has received strong support from HST studies of clusters
at z=0.4-0.5 (Dressler et al. 1997) and ground-based studies at
z=0.1-0.2 (Fasano et al. 2000): these works find the proportion of
spirals to increase and the incidence of S0s to decrease at higher
redshifts.

Based on this, one might expect to observe in cluster S0's some (weak)
residual signs of their past history as spirals.  Perhaps the most
obvious evidence to seek is the presence of relatively young (as
compared to ellipticals) stellar populations.  We have investigated
the ages of stars in ellipticals and S0 galaxies in the Coma cluster
(Poggianti et al. 2001b), deriving luminosity-weighted ages and
metallicities from spectroscopic index-index diagrams (Fig.~1). The
main result is that more than 40\% of the S0s have undergone star
formation during the last $\sim 5$ Gyr, while such activity is absent
in the ellipticals.

A similar conclusion was reached in the Fornax cluster (Kuntschner \&
Davies 1998) and in Abell 2218 (Smail et al. 2001), but other studies
had failed to detect any difference between the ages of Es and S0s. A
possible explanation for the discordant results of different studies
could be the galaxy luminosity range explored. In our Coma sample the
fraction of S0 galaxies with recent star formation is {\sl higher} at
fainter magnitudes, and their quite faint luminosities are consistent
with them being the descendants of typical star-forming spirals at
intermediate redshifts. In contrast, the brightest S0's seem to have
had their latest star formation episode at higher redshifts, and
possibly have formed by some other mechanism, therefore any study
limited to very luminous galaxies might have missed the bulk of the
population with spiral-origins.  The most straightforward way to test
this would be to study the S0 luminosity function in the field and in
the clusters, at low and high-z.

\begin{figure}
\plotone{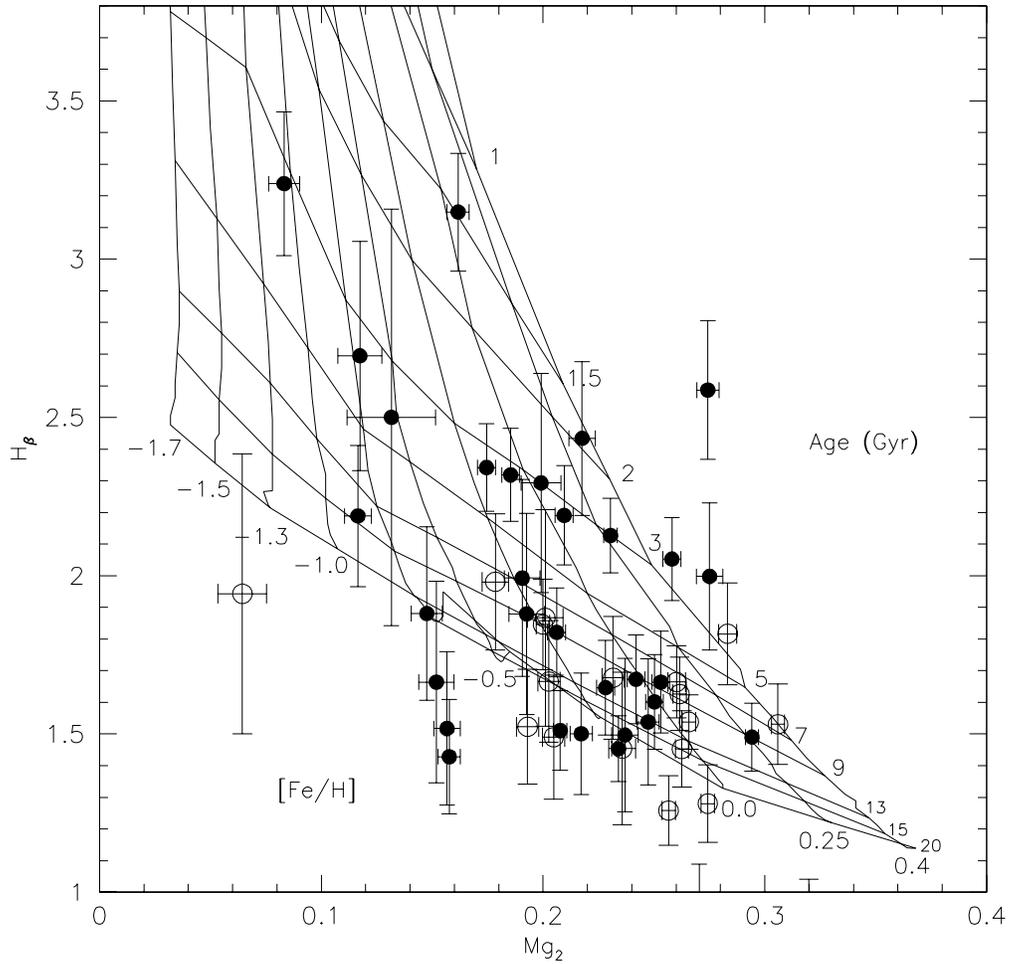}
\caption{$\rm H\beta$ versus $\rm Mg_2$ index strength for ellipticals
(empty circles) and S0 galaxies (filled circles) in Coma 
(Poggianti et al. 2001b). Overplotted are spectrophotometric
models of single stellar populations of different ages and metallicities
as labelled (see paper for details).}
\end{figure}

\section{Galaxy spectra: distant clusters versus Coma}
Additional evidence that starforming galaxies are progressively transformed 
into passive ones in clusters comes from the spectra of distant cluster 
galaxies.

\begin{figure}
\plotone{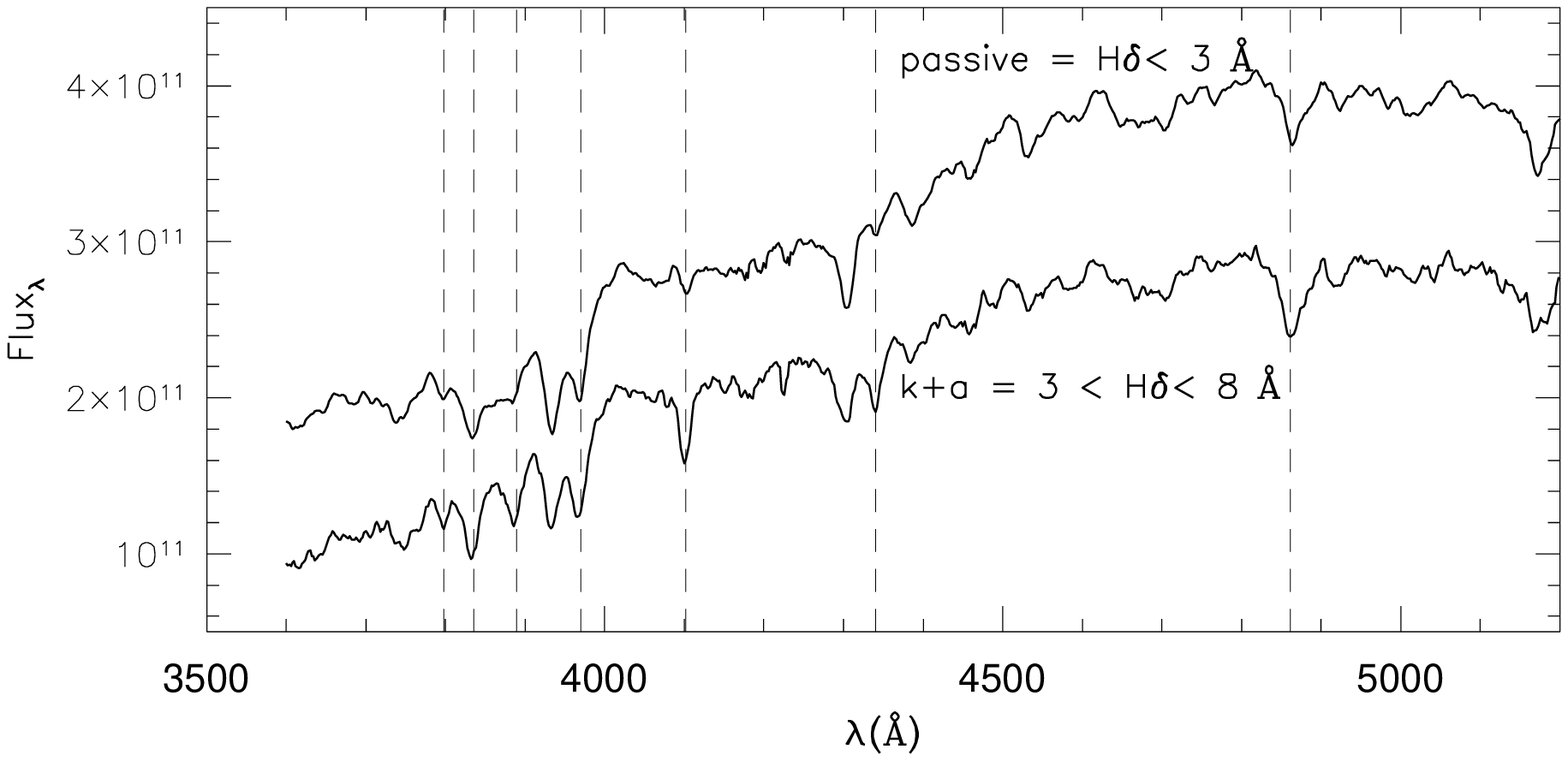}
\caption{Coadded spectra of cluster galaxies at z=0.5: passive galaxies
(top, $\rm EW(H\delta) < 3 $ \AA $\,$) and k+a galaxies (bottom,
$\rm 3 < EW(H\delta) < 8 $ \AA. The Balmer lines (from left
$\rm H\theta$, $\rm H\eta$, $\rm H\zeta$, $\rm H\epsilon$, $\rm H\delta$,
$\rm H\gamma$, $\rm H\beta$) are highlighted by the dashed lines.}
\end{figure}

When the first spectra of distant cluster galaxies were taken
(e.g. Dressler \& Gunn 1983, Couch \& Sharples 1987, Fabricant et al. 1991),
it came as a surprise that many of these spectra displayed unusually strong
Balmer lines in absorption, and no emission lines. This type of spectra,
named ``k+a'' or ``E+A'' spectra, indicate that star formation has recently 
ceased in these galaxies (during the last 1-1.5 Gyr). Those with the
strongest Balmer lines require a starburst before the quenching of star 
formation, and therefore are usually referred to as ``post-starburst 
galaxies''.
The incidence of k+a spectra among cluster galaxies at z=0.4 appears to be much
higher than in field galaxies at similar redshifts (Dressler et al. 1999)
and the majority of k+a spectra belong to galaxies classified as
spirals on the basis of their morphology in HST images (Poggianti et al. 1999).
The differences in the strength of the $\rm H\delta$ and of the other
Balmer lines in k+a spectra as opposed to spectra of passive galaxies
are shown in Fig.~2, where cluster galaxy spectra
at z=0.5 with $\rm 3 < H\delta < 8$ \AA $\,$ and $\rm H\delta < 3$ \AA
$\,$ have been separately coadded (Dressler et al. 2002).

In the following
I am going to show a comparison between the spectral types of distant cluster
galaxies and those of Coma galaxies of similar absolute
magnitudes. The Coma dataset is a spectroscopic survey of a magnitude
limited sample with essentially no additional color or morphological
selection criteria (Mobasher et al. 2001, Poggianti et al. 2002).  The
distant dataset is the spectroscopic catalog of the MORPHS
collaboration (Dressler et al. 1999), that includes 10 rich clusters
at z=0.4-0.5 with a wide range of properties, such as concentration,
optical and X-ray luminosity.  In Table~1 we present the spectral
properties of Coma galaxies, of
the whole MORPHS sample and, separately, of the two
MORPHS clusters whose X-ray luminosities more closely resemble
Coma\footnote{ The X-ray luminosity of Coma (corresponding to $L_X
\sim 9.5 \, 10^{44} \, \rm ergs \, s^{-1}$ when observed at z=0.5 at
0.3-3.5 keV) lies in between that of Cl0016+16 ($L_X \sim 11.8 \,
10^{44} \, \rm ergs \, s^{-1}$, 0.3-3.5 keV) and that of 3C 295 ($L_X
\sim 6.4 \, 10^{44} \, \rm ergs \, s^{-1}$, 0.3-3.5 keV)}.

The most striking difference in Table~1 is the lack of bright k+a
galaxies in Coma.  Moreover, there is a tendency for emission-line
galaxies to be less common in Coma.  Choosing to compare only with
clusters of similar X-ray luminosities enhances the difference between
the k+a population at z=0.5 and at z=0, and attenuates the difference
in the emission-line population.

\begin{table}
\begin{center}
\caption{Comparison between Coma1 and clusters at z=0.5\tablenotemark{a}.\label{tbl-1}}
\begin{tabular}{lrrrr}
\tableline\tableline
Class & \multicolumn{1}{c}{MORPHS\tablenotemark{b}} & Cl 0016+16 & 3C 295 & Coma1 \\
\tableline
passive (k) & 51.1\%  &  56.2\%   & 55.8\%   &  90.6\% \\
k+a/a+k     & 22.4\%  &  32.7\%   & 29.5\%   &  0\%      \\
emission    & 26.5\%  &  11.1\%   & 14.7\%   &  9.4\%   \\
            &&&& \\
N           & 390 & 29 & 25 & 32 \\
\tableline
\tableline
\end{tabular}
\tablenotetext{a}{The data refer to the inner 1.4 Mpc and to $M_V< -19.8$
in all clusters.}
\tablenotetext{b}{All MORPHS clusters.}
\end{center}
\end{table}

\subsection{Emission lines: an interesting but incomplete view}

For distant galaxies the most commonly used indicator of current star
formation activity is the [O{\sc ii}]3727 line, whose flux is expected
to be roughly proportional to the present star formation rate of a
galaxy.

Based on the [O{\sc ii}]3727 line of clusters and field galaxies
in the MORPHS spectroscopic catalog, complemented by a sample
of nearby galaxies (Dressler et al. 2002), we have computed the
SFR per unit B-band luminosity at z=0.5 and z=0, in clusters and in the field,
with the following results:

\vspace{0.3cm}

$z=0.5 \quad \frac{SFR}{L_B}(field) = 3 \times \frac{SFR}{L_B}(clusters) $

\vspace{0.3cm}

$z=0.0 \quad \frac{SFR}{L_B}(field) = 10 \times \frac{SFR}{L_B}(clusters) $

\vspace{0.3cm}

Thus, the difference between the field and the clusters is a factor of 3
more pronounced in the local Universe than at z=0.5.
Furthermore,

\vspace{0.3cm}

${\rm clusters} \quad \frac{SFR}{L_B}(z=0.5) = 6 \times \frac{SFR}{L_B}(z=0.0) $

\vspace{0.3cm}

${\rm field} \qquad \; \; \frac{SFR}{L_B}(z=0.5) = 2 \times \frac{SFR}{L_B}(z=0.0) $

\vspace{0.3cm}

i.e. the evolution of the SFR per unit luminosity, as derived from the
[O{\sc ii}] line, has been steeper in clusters than in the field
during the last 5 Gyrs, as if the cluster environment has somehow
``accelerated'' the evolution towards more passive, earlier-type
galaxies. These results should be always considered remembering that:

a) a lower SFR ([O{\sc ii}] flux) per unit luminosity in the clusters
does not rule out that a SF enhancement could take place in some
or all cluster galaxies {\sl prior} to the final quenching of star 
formation;

b) dust effects are ignored in these calculations, and the amount of
star formation hidden by dust is unknown.  It is important to keep in
mind that starburst galaxies with very high SFR in the local Universe
are characterized by spectra with {\sl weak to moderate} emission
lines, because generally the higher the SFR, the higher the dust extinction
and the fraction of SF hidden.

If dust affected cluster galaxies differently than field galaxies, for
example due to an environmentally-induced star formation enhancement,
then the comparison field-cluster shown above would be misleading.
If, instead, dust effects played the same role in all environments at
a given redshift, the relative comparison field-cluster would be
meaningful.  The evidence of important dust effects in distant cluster
galaxies has been highlighted by the first radio continuum (Smail et
al. 1999) and mid-IR (Duc et al. 2002) studies.

\section{Fainter galaxies}
Any comparison with the distant cluster data is inevitably limited to
the brightest end of the galaxy luminosity function, while in nearby
clusters it is possible to study also fainter galaxies.  In Coma, our
spectroscopic survey extends over more than 6 magnitudes down to $M_B
\sim -14$.

In the previous section I have discussed the fact that Coma lacks the
population of {\sl luminous} k+a galaxies that is present in clusters
at z=0.4. Strikingly, very clear examples of k+a spectra are found
instead among the {\sl faint} Coma galaxies, as shown in Fig.~3
(Poggianti et al. 2002).  We show here a sample of k+a's with a range
of $\rm H\delta$ strengths, from very strong, post-starburst galaxies
(\#1 and \#4) to weaker cases (\#6 and \#7). Balmer-strong spectra
of Coma galaxies have been previously reported in a number of works
by Caldwell, Rose and collaborators (e.g. Rose et al. 2001 and references
therein).

Coma k+a's in our sample are typically fainter than $M_V \sim -18.5$
(Fig.~4) and constitute a significant fraction (10-15 \%) of the dwarf
galaxy population at $M_B > -17$.  Why are k+a's in clusters a {\sl
luminous} phenomenon at z=0.5, and a {\sl faint} phenomenon at z=0? 
Is this due to a ``cosmic evolution'' (i.e. the changes in field galaxies +
the hierarchical history of clusters), or to an evolution with z of
the {\sl processes} affecting galaxies in clusters? 
The former possibility is probably more likely. In fact, when studying the 
luminosity-weighted ages of all Coma galaxies with no emission lines
in our sample we find systematic trends of age with galaxy luminosity.
Fig.~5 shows what proportion of currently-passive Coma galaxies of a 
given magnitude experienced the latest star formation during the last 3 Gyr, 
between 3 and 9 Gyr ago and more than 9 Gyr ago. The age of the
latest star formation episode was derived from index-index diagrams
similar to Fig.~1, as described in Poggianti et al. 2001a.

About half of all galaxies of any luminosity do not show signs of significant
star formation activity in the last 9 Gyr, i.e. since z=1.4 in a
$\Lambda$ cosmology. Among the bright, currently passive galaxies, 
the fraction that experienced a star formation
activity betwen 3 and 9 Gyr ago ($0.25 < z < 1.4$) is much higher than
the fraction with SF during the last 3 Gyr (since $z=0.25$), while
the opposite is true for the faint galaxy population.
This seems to point to a ``downsizing effect'' suggesting that
the last star formation activity (possibly related to the epoch
of accretion onto the cluster) occurs on average at lower redshifts for 
progressively fainter galaxies. In this scenario, the galactic properties 
become a probe of the cluster accretion history of field starforming
galaxies.

\begin{figure}
\plotone{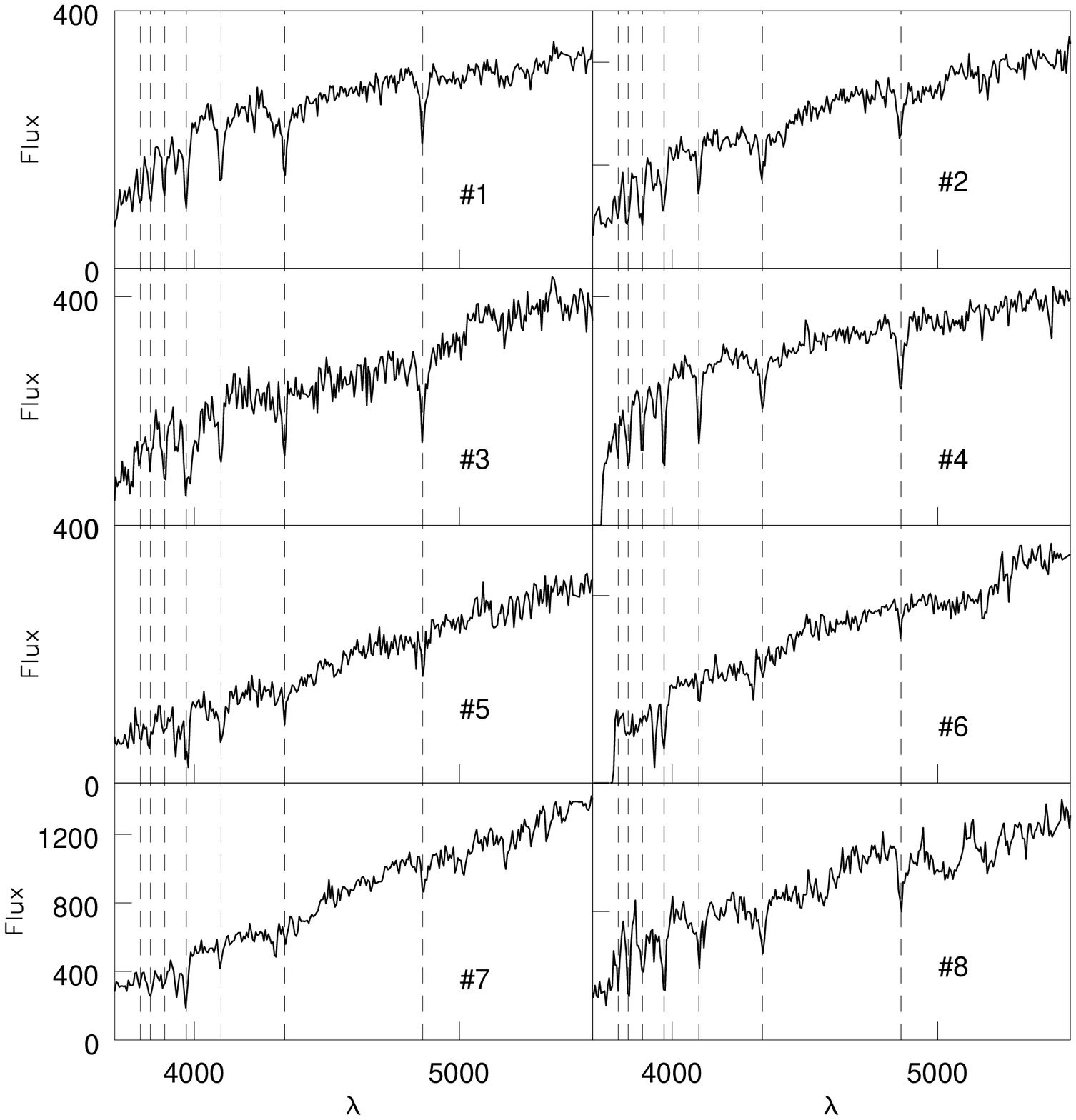}
\caption{Spectra of faint k+a galaxies in Coma. The Balmer lines (from left
$\rm H\theta$, $\rm H\eta$, $\rm H\zeta$, $\rm H\epsilon$, $\rm H\delta$,
$\rm H\gamma$, $\rm H\beta$) are highlighted by the dashed lines. 
}
\end{figure}

\begin{figure}
\plotone{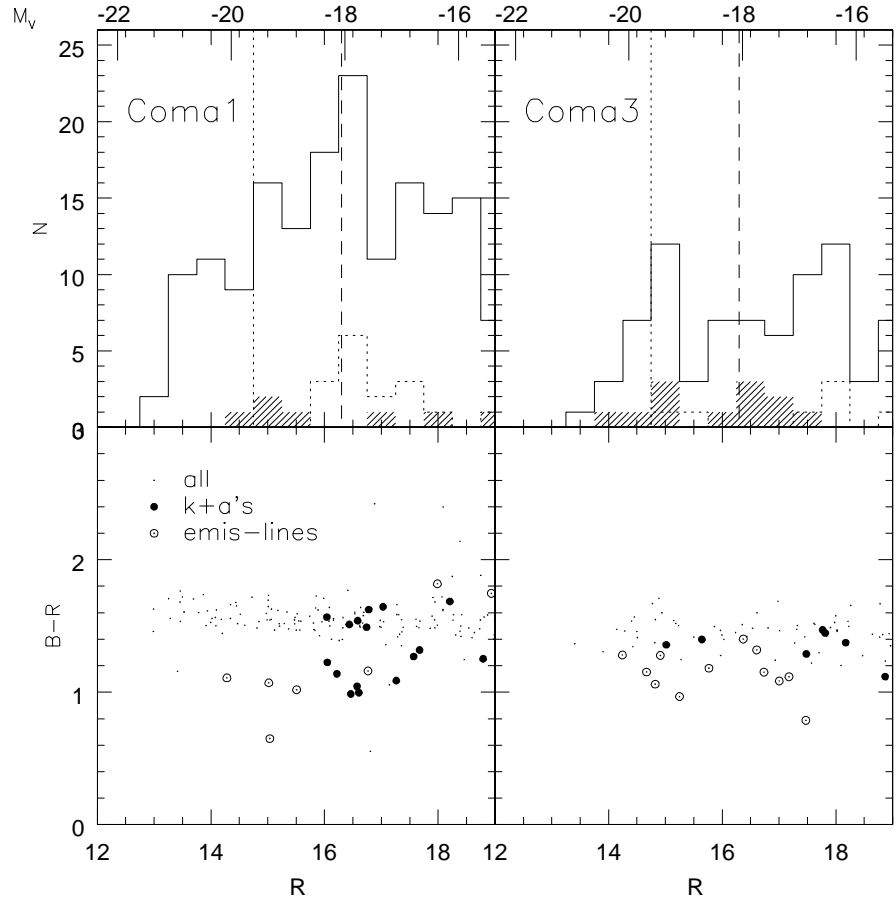}
\caption{Magnitude distribution (top) and color-magnitude diagram (bottom)
of galaxies in the center (Coma1) and South-West region (Coma3)
of Coma. The dotted line
shows the magnitude limit corresponding to the MORPHS limit. The dashed
line is the adopted magnitude division between ``giants'' and ``dwarfs''.
Total histogram= all. Dashed histogram = k+a's. Shaded histogram =
emission-line galaxies. The corresponding absolute V band magnitude 
scale is shown on top of the plot.}
\end{figure}

\begin{figure}
\plotone{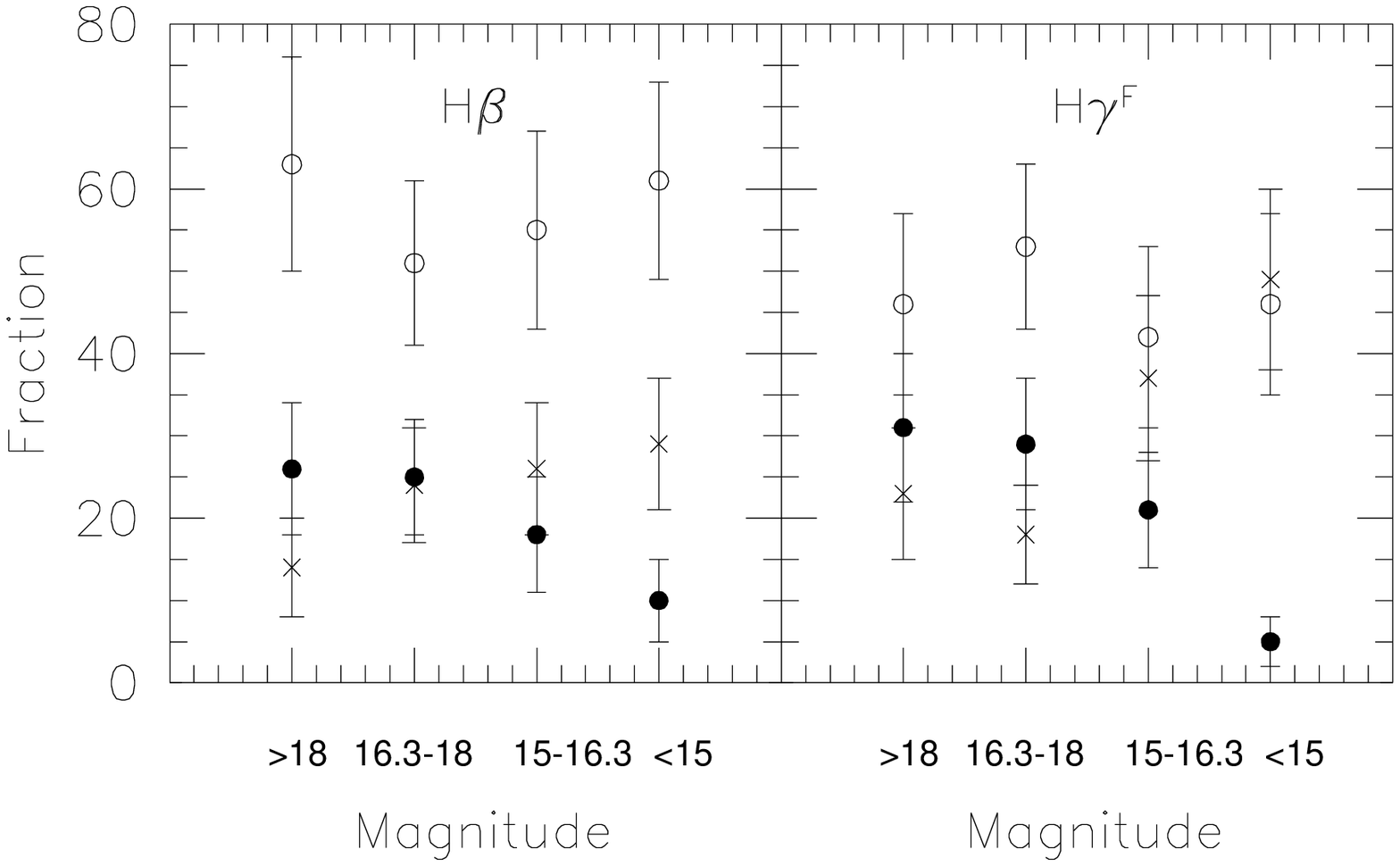}
\caption{Fraction of young (filled dots, age $<3$ Gyr), 
intermediate-age (crosses, 3 to 9 Gyr) and old (empty dots, age $>9$ Gyr) 
Coma1 galaxies within each R magnitude bin as
derived from the $\rm H\beta$/$\rm Mg_2$ diagram (left) and the 
$\rm H_{\gamma}^F$/$<$Fe$>$ diagram (right). The errorbars
are Poissonian.}
\end{figure}





\acknowledgments
I wish to thank my collaborators of the Coma group, in particular T. Bridges, B. Mobasher and D. Carter, and of the MORPHS group, especially A. Dressler
and A. Oemler, for countless useful discussions and for allowing me to show
our common results ahead of publication.

\end{document}